\documentclass[12pt,prb,aps,amssymbols]{revtex4}
\usepackage{amssymb}
\usepackage{bm}
\usepackage{amsbsy}
\usepackage{epsfig}
\usepackage{color}
\usepackage{amsmath}
\begin{document}

\date{\today}

\title{ \begin{center}
Nonlinear parallel momentum transport in strong turbulence
\end{center}}

\author{Lu Wang,$^{1}$ Tiliang Wen,$^1$ and P. H. Diamond$^{2}$\\
$^1$State Key Laboratory of Advanced Electromagnetic Engineering and Technology, School of Electrical and Electronic Engneering, Huazhong University of Science and Technology, Wuhan 430074, China\\
$^2$Center for Momentum Transport and Flow Organization and\\Center for Astrophysics and Space Sciences,\\ University of California at San Diego, La Jolla, CA 92093-0424, USA\\
E-mail: luwang@hust.edu.cn}

\maketitle \maketitle

\section*{Abstract}

Most existing theoretical studies of momentum transport focus on calculating the Reynolds stress based on quasilinear theory, without considering the \emph{nonlinear} momentum flux-$\langle \tilde{v}_r \tilde{n} \tilde{u}_{\|} \rangle$. However, a recent experiment on TORPEX found that the nonlinear toroidal momentum flux induced by blobs makes a significant contribution as compared to the Reynolds stress [Labit et al., Phys. Plasmas {\bf 18}, 032308 (2011)]. In this work, the nonlinear parallel momentum flux in strong turbulence is calculated by using three dimensional Hasegawa-Mima equation. It is shown that nonlinear diffusivity is smaller than quasilinear diffusivity from Reynolds stress. However, the leading order nonlinear residual stress can be comparable to the quasilinear residual stress, and so could be important to intrinsic rotation in tokamak edge plasmas. A key difference from the quasilinear residual stress is that parallel fluctuation spectrum asymmetry is not required for nonlinear residual stress.

\section{Introduction}

Tokamak plasma rotation and toroidal angular momentum transport have been subjects of intensive study due to their important role in reducing turbulent transport as well as in stabilizing magnetohydrodynamic (MHD) instability, such as resistive wall modes\cite{A. Bondeson and D. J. Ward, R. Betti and J. P. Freidberg}. Owing to a wide range of beneficial effects on stability, confinement, and performance of tokamak plasmas, much effort has been devoted to understanding the mechanisms underlying the change in rotation and how to control it. On one hand, there are a number of known causes for the plasma rotation to slow down, such as nonaxisymmetric error fields,\cite{S. M. Wolfe,R. Fitzpatrick} loss of momentum input as a consequence of Alfven eigenmodes,\cite{E. J. Strait} and edge localized modes (ELMs).\cite{B Goncalves} On the other hand, toroidal rotation is driven externally by neutral beam injection (NBI) on most present day devices. However, beam injection may be of limited utility in providing enough external torque in future reactor such as International Thermonuclear Experimental Reactor. One alternative is to take advantage of intrinsic rotation (spontaneous, or self-generated, in the absence of an external momentum input) which has been widely observed under a variety of operating conditions.\cite{Rice08} Consequently, understanding plasma rotation and momentum transport under low external momentum input condition is of major interest.

The total flux of parallel momentum driven by electrostatic turbulence has the form\cite{Diamond09}
\begin{equation}\label{momentum flux}
\Pi_{r,\|} = \langle n \rangle \langle \tilde{v}_r \tilde{u}_{\|} \rangle + \langle U_{\|} \rangle \langle \tilde{v}_{r} \tilde{n} \rangle + \langle \tilde{v}_r \tilde{n} \tilde{u}_{\|} \rangle.
\end{equation}
Here, on the right hand side (RHS) of Eq.~(\ref{momentum flux}), the first term is the parallel Reynolds stress, the second term is convection, due to particle flux, and the third term is the \emph{nonlinear} flux. The Reynolds stress can be further decomposed as
\begin{equation}\label{Reynolds stress}
\langle \tilde{v}_r \tilde{u}_{\|} \rangle = - \chi_{\varphi} \frac{\partial \langle U_{\|} \rangle}{\partial r} + V \langle U_{\|} \rangle + \Pi_{r,\|}^{Res},
\end{equation}
where, on the RHS, they are diffusion, pinch term and residual stress, respectively. The residual stress is thought to be the origin of the intrinsic rotation which has been intensively investigated by using quasilinear theory.\cite{Diamond2013} In addition, turbulent acceleration is proposed as another possible mechanism for driving intrinsic rotation.\cite{Lu Wang PRL2013,Garbet PoP2013} Turbulent acceleration acts as a local source or sink, which has different physics from the residual stress. The third term on the RHS of Eq.~(\ref{momentum flux}), $\langle \tilde{v}_r \tilde{n} \tilde{u}_{\|} \rangle$, represents the nonlinear (as opposed to quasilinear) flux, driven by processes such as mode-mode coupling and turbulence spreading.\cite{Diamond09} Most existing theoretical works on parallel momentum transport neglect the nonlinear flux, which is less understood.\cite{Diamond2013} However, the nonlinear flux may also influence the rotation profile, especially at the boundary where the relative fluctuation amplitude can be strong, i.e., since $ \displaystyle \frac{\tilde{n}}{n_0}\rightarrow 1 $, the nonlinear flux cannot be dismissed as small. In this sense, momentum transport theory is still not well developed.

Recent experimental results on TORPEX showed that blob induced fluctuations are so strong that the toroidal flow is transiently reversed, and the associated nonlinear toroidal momentum flux can be dominant for some time.\cite{Labit11} This result suggests that the nonlinear flux is no longer negligible. Similar blobs in L-mode\cite{Terry,Zweben,Grulke,Agositni,Xu09,Chen10,Maqueda} and edge localized mode (ELM) filaments in H-mode\cite{Rudakov,Boedo03,Fundamenski,Endler,Boedo,Kirk,Herrmann,Schmid,Dudson,Silva} are observed in tokamaks. In general, the nonlinear flux is of potential relevance in the strongly turbulent edge. Therefore, to fully comprehend momentum transport, a theoretical study of the nonlinear momentum flux in strong turbulence seems necessary.

In present work, we calculate the nonlinear parallel momentum flux by using the three dimensional Hasegawa-Mima equation\cite{Hasegawa and Mima} containing the compression of ion parallel velocity and the ion parallel momentum equation. For comparison, the parallel Reynolds stress is also calculated. We find that the nonlinear diffusivity is small compared to the quasilinear diffusivity from the Reynolds stress. However, the dominant nonlinear residual stress can be comparable to the quasilnear residual stress with opposite sign, if increasing fluctuation intensity profile is used for the residual stress.\cite{Gurcan2010} This indicates that strong momentum transport induced by blob ejection at edge is important to intrinsic rotation. We also find that parallel fluctuation spectrum asymmetry is not necessary for nonlinear residual stress, in contrast to the case of quasilinear residual stress.

The remainder of this paper is organized as follows. Sec.~II presents the minimal model adopted in this work. The nonlinear momentum flux and its comparison to the Reynolds stress are presented in Sec.~III. Finally, we summarize our work and discuss the implications for momentum transport and rotation response to blob ejection in Sec.~IV. In the appendix, we present details of the calculation.

\section{Minimal theoretical model}

To obtain the triple nonlinear momentum flux, $\langle \tilde{v}_r \tilde{n} \tilde{u}_{\|} \rangle$, we need to calculate the coherent part of fluctuations for the beat mode, and then use the two-scale direct interaction approximation (TSDIA).\cite{Gurcan2006, Diamond book} In this way, the nonlinear parallel momentum flux can be written as
\begin{equation}\label{NLflux}
\Pi_{r,\|}^{NL} = \frac{1}{3} \left( \langle \tilde{v}_r^{(c)} \tilde{n} \tilde{u}_{\|} \rangle + \langle \tilde{v}_r \tilde{n}^{(c)} \tilde{u}_{\|} \rangle + \langle \tilde{v}_r \tilde{n} \tilde{u}_{\|}^{(c)} \rangle \right).
\end{equation}
Here, the subscript (c) means the coherent component of the beat mode. $ \tilde{v}_r=-ik_y \displaystyle \frac{c}{B}\tilde{\phi} $ is the radial fluctuating $ E\times B $ drift velocity. For simplicity, the adiabatic approximation$ \displaystyle \frac{\tilde{n}}{n_0} = \frac{e\tilde{\phi}}{T_e} $ is used, so we have $ \displaystyle \frac{\tilde{n}^{(c)}}{n_0} = \frac{e\tilde{\phi}^{(c)}}{T_e} $. Thus, we take mode widths large enough so that $k_{\|}^{\prime2}\triangle^2 v_{the}^2 /(\omega_k \nu_{ei}) >1$, with $k_{\|}^{\prime}$ the radial derivative of parallel wave number, $\triangle$ the mode width, $v_{the}$ the electron thermal velocity, and $\nu_{ei}$ the electron-ion collision frequency. $k_{\|}=0$ driven mode effects are neglected here. Now, the coherent parts of $ \tilde{\phi} $ and $ \tilde{u}_{\|} $ are required.
In this work, we adopt three dimensional Hasegawa-Mima (H-M) equation with parallel flow compression which can be written as
\begin{eqnarray}\label{H-M}
\frac{\partial}{\partial t}\left(\rho_s^2 \nabla_{\perp}^2\phi - \phi \right) + \rho_s^4 \omega_{ci} \hat{z} \times \nabla \phi \cdot \nabla \nabla_{\perp}^2\phi -i\omega_{*n} \phi&=& c_s\nabla_{\|} u_{\|},
\end{eqnarray}
and the parallel momentum equation for cold ions
\begin{eqnarray}
\left(\frac{\partial}{\partial t} + \omega_{ci} \rho_s^2 \hat{z} \times \nabla \phi \cdot \nabla \right)u_{\|}  - \rho_s \frac{\partial}{\partial r}\langle U_{\|} \rangle  \frac{\partial}{\partial y}\phi &= - c_s \nabla_{\|} \phi \label{para velocity}.
\end{eqnarray}
Here, we have used the standard normalization for electric potential fluctuation $\phi \equiv e \tilde{\phi} / T_e$, parallel velocity fluctuation $u_{\|} \equiv \tilde{u}_{\|} / c_s$,  with $\omega_{ci}= eB/(m_i c)$ the ion gyrofrequency, $c_s$ the ion acoustic velocity, and $\rho_s=\frac{c_s}{\omega_{ci}}$ the ion Larmor radius at the electron temperature. For the spatial scale, we consider two-scale approach, i.e., $\nabla_{\perp} = i{\bf k}_{\perp}+  {\partial}/{\partial r}$, where ${\bf k}_{\perp}$ denotes wave number of the fast spatial fluctuations, and $\partial /\partial r$ describes modulation of the wave envelope, which occurs on a slowly varying spatial scale. $\omega_{*n}=k_y \rho_s c_s / L_n$ is the electron diamagnetic drift frequency with $L_n = -\left(\partial \ln n /\partial r\right)^{-1}$ density gradient scale length, and $\langle U_{\|}\rangle$ is the mean parallel flow velocity. The last term on the RHS of Eq.~(\ref{H-M}) comes from ion parallel compression. In Eq.~(\ref{para velocity}), the assumptions of isothermal electrons and $\omega_k \gg k_{\|} \langle U_{\|}\rangle$ are used, and ion pressure gradient, $\nabla_{\|} P_i$, is absent due to the cold ion approximation.


Taking the Fourier transformations of Eqs.~(\ref{H-M})-(\ref{para velocity}) yields
\begin{subequations}
\begin{align}
\frac{\partial}{\partial t} \phi_k + i \frac{\omega_{*n}}{1+k_{\perp}^2{\rho_s^2}} \phi_k + \frac{ik_{\|}c_s}{1+k_{\perp}^2{\rho_s^2}} u_k &= \sum_{k=k^{\prime} + k^{\prime\prime}} M^1_{k,k^{\prime},k^{\prime\prime}},\label{H-Mk}\\
\frac{\partial}{\partial t} u_k -ik_y\rho_s \frac{\partial}{\partial r} \langle U_{\|} \rangle \phi_k +ik_{\|} c_s \phi_k &= \sum_{k=k^{\prime} + k^{\prime\prime}} M^2_{k,k^{\prime},k^{\prime\prime}}\label{uk},
\end{align}
\end{subequations}
where the nonlinear terms are
\begin{subequations}
\begin{align}
M^{1}_{k,k^{\prime},k^{\prime\prime}} = & \frac{\omega_{ci}}{2 ( 1 + k_{\perp}^2 {\rho_s^2})} {\rho_s^4} \left\lbrace\hat{z}\times {\bf k}_{\perp}^{\prime} \cdot {\bf k}_{\perp}^{\prime\prime}(k_{\perp}^{\prime\prime2} - k_{\perp}^{\prime2})\phi_{k^{\prime}}\phi_{k^{\prime\prime}}\right.\nonumber\\
& \left. +i\phi_{k^{\prime}} \frac{\partial}{\partial r} \phi_{k^{\prime\prime}} \left[k_y^{\prime}(k_{\perp}^{\prime\prime2} - k_{\perp}^{\prime2}) - 2k_x^{\prime\prime}\hat{z}\times {\bf k}_{\perp}^{\prime} \cdot {\bf k}_{\perp}^{\prime\prime}\right]\right.\nonumber\\
& \left. - i\phi_{k^{\prime\prime}} \frac{\partial}{\partial r} \phi_{k^{\prime}} \left[k_y^{\prime\prime}(k_{\perp}^{\prime\prime2} - k_{\perp}^{\prime2}) - 2k_x^{\prime}\hat{z}\times {\bf k}_{\perp}^{\prime} \cdot {\bf k}_{\perp}^{\prime\prime}\right]
\right\},\\
M^{2}_{k,k^{\prime},k^{\prime\prime}} = & \frac{\omega_{ci}}{2}  \hat{z}\times {\bf k}_{\perp}^{\prime} \cdot {\bf k}_{\perp}^{\prime\prime} {\rho_s^2} \left(\phi_{k^{\prime}}u_{k^{\prime\prime}} - u_{k^{\prime}}\phi_{k^{\prime\prime}}\right)\nonumber\\
&+ \frac{\omega_{ci}}{2}  ik_y^{\prime} {\rho_s^2} \left(\phi_{k^{\prime}}\frac{\partial}{\partial r}u_{k^{\prime\prime}} - u_{k^{\prime}} \frac{\partial}{\partial r} \phi_{k^{\prime\prime}}\right)- \frac{\omega_{ci}}{2}  ik_y^{\prime\prime} {\rho_s^2} \left(u_{k^{\prime\prime}} \frac{\partial}{\partial r} \phi_{k^{\prime}} - \phi_{k^{\prime\prime}}\frac{\partial}{\partial r}u_{k^{\prime}} \right).
\end{align}
\end{subequations}
Here, the higher order terms related to slow spatial variation $\displaystyle \frac{\partial ^2}{\partial r^2}$ have been neglected. Eqs.~(\ref{H-Mk}) and (\ref{uk}) can be expressed compactly in the form of a matrix as follows.
\begin{equation}\label{etak}
\frac{\partial}{\partial t} \eta_k^{\alpha} + H_k^{\alpha \beta} \eta_k^{\beta} = \sum_{k=k^{\prime} + k^{\prime\prime}} M_{k,k^{\prime},k^{\prime\prime}}^{\alpha},
\end{equation}
with
\begin{equation}
\eta_k=\left[
  \begin{array}{c}
    \phi_k \\
   u_k \\
  \end{array}
\right], \nonumber
\end{equation}
\begin{equation}
H=\left[
  \begin{array}{cc}
   \displaystyle  i \frac{\omega_{*n}}{1+k_{\perp}^2{\rho_s^2}} & \displaystyle \frac{ik_{\|}c_s}{1+k_{\perp}^2{\rho_s^2}} \\
  \displaystyle ik_{\|}c_s - ik_y \rho_s \frac{\partial}{\partial r} \langle U_{\|} \rangle & 0 \\
  \end{array}
\right] \equiv \left[
  \begin{array}{cc}
   a & b \\
  c & 0 \\
  \end{array}
\right] ,\nonumber
\end{equation}
and
\begin{equation}
\sum_{k=k^{\prime} + k^{\prime\prime}} M_{k,k^{\prime},k^{\prime\prime}} = \left[
  \begin{array}{c}
    \displaystyle \sum_{k=k^{\prime} + k^{\prime\prime}} M^1_{k,k^{\prime},k^{\prime\prime}} \\
   \displaystyle \sum_{k=k^{\prime} + k^{\prime\prime}} M^2_{k,k^{\prime},k^{\prime\prime}} \\
  \end{array}
\right].\nonumber
\end{equation}

The linear theory of this three dimensional Hasegawa-Mima system is clear. The dispersion equation is as follows:
\begin{subequations}
\begin{align}
\lambda_1 = i\omega_{k1} \approx \frac{i \omega_{*n}}{1+k_{\perp}^2 \rho_s^2} + i \frac{k_{\|}^2 c_s^2 }{\omega_{*n}}, \label{omega_1} \\
\lambda_2 = i\omega_{k2} \approx - i \frac{k_{\|}^2 c_s^2 }{\omega_{*n}} . \label{omega_2}
\end{align}
\end{subequations}

The nonlinear terms are crucial to produce the coherent parts of the beat mode. By using the Markovian approximation, the nonlinear coupling terms can be written as
\begin{equation}
\sum_{k=k^{\prime} + k^{\prime\prime}} M_{k,k^{\prime},k^{\prime\prime}}^{\alpha} = - \gamma_{k.\alpha}^{NL} \eta_k^{\alpha} + F_{k,\alpha} + 2M_{k,k^{\prime},k^{\prime\prime}}^{\alpha}, \label{Markovian approximation}
\end{equation}
where $\gamma_{k.\alpha}^{NL}$ is the eddy-damping rate, and $F_{k,\alpha}$ is fast fluctuating force which does not contribute to coherent parts of the beat mode. \emph{Note that the nonlinear damping rate is larger than the frequency mismatch for strong turbulence}. By diagonalization of the matrix $H$ in Eq.~(\ref{etak}), the coherent component of beat mode can be obtained as follows:
\begin{equation}\label{coherent}
\eta_k^{\alpha(c)}(t)= \int_{-\infty} ^t dt^{\prime} R_k^{\alpha \beta}(t,t^{\prime}) M_{k,k^{\prime},k^{\prime\prime}}^\beta,
\end{equation}
where the response function $R_k^{\alpha \beta}(t, t^{'})$ is
\begin{equation}\label{response function}
R_k^{\alpha \beta}(t, t^{'})= r_k^{\alpha \beta \delta} exp \left[ \left( i\omega_{k \delta} + \gamma_{k \delta}^{NL} \right) (t^{\prime}-t) \right],
\end{equation}
with
\begin{equation}
r_k^{\alpha \beta 1} \cong \left[
  \begin{array}{cc}
   1 & \displaystyle \frac{k_{\|}c_s} {\omega_{*n}} \\
  \displaystyle \frac{ k_{\|}c_s(1+k_{\perp}^2 \rho_s^2)} {\omega_{*n}} & \displaystyle \frac{ k_{\|}^2 c_s^2 (1+k_{\perp}^2 \rho_s^2) } {\omega_{*n}^2} \\
  \end{array}
\right] , \nonumber
\end{equation}
\begin{equation}
r_k^{\alpha \beta 2} \cong \left[
  \begin{array}{cc}
  \displaystyle \frac{ k_{\|}^2 c_s^2 (1+k_{\perp}^2 \rho_s^2) } {\omega_{*n}^2}  & \displaystyle \frac{-k_{\|}c_s} {\omega_{*n}} \\
  \displaystyle \frac{- k_{\|}c_s(1+k_{\perp}^2 \rho_s^2)} {\omega_{*n}}  & 1 \\
  \end{array}
\right]. \nonumber
\end{equation}
The details of calculation are presented in the Appendix A.

Inserting the coherent component, Eq.~(\ref{coherent}) into the nonlinear momentum flux, Eq.~(\ref{NLflux}), then we need to calculate the forth order moment terms. By using the approximation of quasi-Gaussian statistics, the forth order moment can be decoupled into a product of quadratic moments, i.e.,
\begin{equation}
\langle \eta_{k^{\prime}}(t) \eta_{k^{\prime}}^*(t^{\prime}) \eta_{k^{\prime\prime}}(t) \eta_{k^{\prime\prime}}^*(t^{\prime})\rangle = \langle \eta_{k^{\prime}}(t) \eta_{k^{\prime}}^*(t^{\prime})\rangle \langle  \eta_{k^{\prime\prime}}(t) \eta_{k^{\prime\prime}}^*(t^{\prime}) \rangle.
\end{equation}
Here, within the Markovian approximation, the quadratic moments, or in another word, the two-time correlation function can be expressed by one-time correlation functions as
\begin{equation}
 \langle \eta_k ^{\alpha *}(t^{\prime}) \eta_k^{\beta}(t) \rangle = exp [i\omega_k (t^{\prime}-t)-\gamma_{k,\alpha}^{NL} |t^{\prime}-t| ] \langle \eta_k ^{\alpha *}(t) \eta_k^{\beta}(t) \rangle.
\end{equation}
Now, we have all the essentials for evaluation of the nonlinear momentum flux.

\section{Nonlinear residual stress and comparison}

In this section, we present the results for nonlinear momentum flux without showing the tedious calculations. The details of calculation can be found in Appendix B. Here, we write the nonlinear parallel momentum flux again as follows:
\begin{equation}\label{NLfluxnew}
\Pi_{r,\|}^{NL} = \frac{1}{3} \left( \langle \tilde{v}_r^{(c)} \tilde{n} \tilde{u}_{\|} \rangle + \langle \tilde{v}_r \tilde{n}^{(c)} \tilde{u}_{\|} \rangle + \langle \tilde{v}_r \tilde{n} \tilde{u}_{\|}^{(c)} \rangle \right).
\end{equation}
We need to substitute the coherent components $\phi_k^{(c)}$ into the first two terms and $u_k^{(c)}$ into the last term on RHS of Eq.~(\ref{NLfluxnew}) to calculate the nonlinear momentum flux. The results of the first two nonlinear momentum flux terms can be written as
\begin{eqnarray}
\Pi_{r,\|1}^{NL} &=& \langle \tilde{v}_r^{(c)} \tilde{n} \tilde{u}_{\|} \rangle + \langle \tilde{v}_r  \tilde{n}^{(c)} \tilde{u}_{\|} \rangle \nonumber\\
 &=& n_0 c_s^2 \Re\sum_{k=k^{\prime} + k^{\prime\prime}}\left[\left(ik_y-ik_y^{\prime} \right)\rho_s \int_{-\infty}^{t}dt^{\prime} \langle R_k^{1\beta*}(t,t^{\prime}) M_{k,k^{\prime},k^{\prime\prime}}^{\beta*}(t^{\prime}) n_{k^{\prime}}(t) u_{k^{\prime\prime}}(t)\rangle \right] \nonumber\\
 &=& - n_0 \chi_1^{NL}\frac{\partial \langle U_{\|} \rangle }{\partial r} + n_0 \Pi_{r,\|1}^{NL,res},
\end{eqnarray}
with the leading order nonlinear diffusivity is
\begin{eqnarray}
 \chi_1^{NL}&=&  \frac{1}{4}\rho_s c_s \sum_{k=k^{\prime} + k^{\prime\prime}} \frac{\tau_{c1} \omega_{ci} }{ 1 + k_{\perp} ^2 \rho_s^2} I_{k^{\prime}}I_{k^{\prime\prime}} \frac{\rho_s}{L_I} g_{k^{\prime\prime}} A_{k^{\prime},k^{\prime\prime}} \nonumber,
\end{eqnarray}
and the leading order nonlinear residual stress is
\begin{eqnarray}
\Pi_{r,\| 1}^{NL,res}&=&-\frac{1}{4}n_0 c_s^2 \sum_{k=k^{\prime}+k^{\prime\prime}} I_{k\prime}I_{k\prime\prime} \left[\frac{\tau_{c1}\omega_{ci}}{(1+k_{\perp}^2 \rho_s^2)} g_{k\prime\prime}A_{k\prime,k\prime\prime}\frac{\Delta^{\prime\prime2}}{L_sL_I} \frac{\rho_s}{L_I}+2\frac{L_n}{L_s}        \tau_{c2}\omega_{ci} k_y^{\prime\prime2}\rho_s^2g_{k\prime}g_{k\prime\prime}\frac{\Delta^{\prime\prime2}}{L_s^2} \right]. \nonumber
\end{eqnarray}
Here, $\tau_{c1}$ and $\tau_{c2}$ are triad interaction time for vorticity equation and parallel momentum equation, respectively. They can be estimated by the inverse of corresponding nonlinear damping rates, because the nonlinear damping rate is much larger than the frequency mismatch in strong turbulence. $I_k = |\phi_k|^2$ is the fluctuation intensity, $ \displaystyle\frac{1}{L_I} = \displaystyle \frac{1}{I_k}\frac{\partial }{\partial r}I_k $ is the intensity gradient scale length, $L_s$ is the magnetic shear scale length,  $\triangle$ is the mode width, and other dimensionless parameters are $A_{k^{\prime},k^{\prime\prime}} = k_y^{\prime\prime2} \left(k_{\perp}^{\prime2} - k_{\perp}^{\prime\prime2} + 2k_{x}^{\prime2}\right)\rho_s^4 $, $ \displaystyle g_k = \frac{k_yc_s \omega_k}{\left(\omega_k^2 + \gamma_{k,NL2}^2\right)} $. Note that $ L_I $ is positive for increasing intensity from inside to outside, and $ L_s $ is positive for normal magnetic shear. $ A_{k^{\prime},k^{\prime\prime}}>0 $ for $ k_{\perp}^{\prime}\sim k_{\perp}^{\prime\prime} $, and $ g_k $ is always positive. Therefore, the leading order nonlinear diffusivity satisfies $ \chi_1^{NL}>0  $ for increasing intensity profile in the edge regime. The sign of $ \Pi_{r,\| 1}^{NL,res} $ depends on the sign of $ L_s $, and so is negative for normal magnetic shear.


The other nonlinear momentum flux term can be written as
\begin{eqnarray} \label{Pi_2leading}
\Pi_{r,\|2}^{NL} &=& \langle \tilde{v}_r \tilde{n} \tilde{u}_{\|}^{(c)} \rangle  \nonumber\\
&=& n_0 c_s^2\ \Re\sum_{k=k^{\prime} + k^{\prime\prime}}-ik_y^{\prime}\rho_s \int_{-\infty}^{t}dt^{\prime} \langle R_k^{2 \beta*}(t,t^{\prime}) M_{k,k^{\prime},k^{\prime\prime}}^{\beta*}(t^{\prime})  \phi_{k^{\prime}}(t) n_{k^{\prime\prime}}(t)\rangle \nonumber\\
 &=& - n_0 \chi_2^{NL}\frac{\partial \langle U_{\|} \rangle }{\partial r} + n_0 \Pi_{r,\|2}^{NL,res},
\end{eqnarray}
where the leading order nonlinear diffusivity is
\begin{eqnarray}
 \chi_2^{NL}&= & \frac{1}{4} \rho_s c_s \sum_{k=k^{\prime} + k^{\prime\prime}}\tau_{c2} \omega_{ci} I_{k^{\prime}}I_{k^{\prime\prime}} \left(g_{k^{\prime}} -g_{k^{\prime\prime}} \right)k_y^{\prime2}\rho_s^2 \frac{\rho_s}{L_I} \nonumber,
\end{eqnarray}
and the leading order nonlinear residual stress is
\begin{eqnarray}
\Pi_{r,\|2}^{NL,res}&= & \frac{1}{2} c_s^2 \sum_{k=k^{\prime} + k^{\prime\prime}}\tau_{c2} \omega_{ci} I_{k^{\prime}}I_{k^{\prime\prime}} g_{k^{\prime\prime}} k_y^{\prime2} \rho_s^2 \frac{\rho_s}{L_s} \nonumber.
\end{eqnarray}
Here, the sign of $ \chi_2^{NL}$ is not clear. We can rewrite its expression in terms of symmetric $ k^{\prime} $ and $ k^{\prime\prime} $, $ \chi_2^{NL}\propto \left( g_{k^{\prime}} -g_{k^{\prime\prime}} \right) \left( k_y^{\prime2}-k_y^{\prime\prime2} \right) $ , which is positive. Then, one can find that the sign of $ \chi_2^{NL}$ is the same as that of $ \chi_1^{NL}$, i.e., it is positive for increasing intensity profile. However, the sign of nonlinear residual stress, $ \Pi_{r,\|2}^{NL,res} $ is opposite to that of $ \Pi_{r,\|1}^{NL,res} $, i.e., $ \Pi_{r,\|2}^{NL,res} $ is positive for normal magnetic shear.

To campare with the usual Reynolds stress, we also calculate it quasilinearlly.
\begin{eqnarray}
\displaystyle \Pi_{r,\|}^{Rey} &=& n_0 \langle \tilde{v}_r \tilde{u}_{\|} \rangle \nonumber\\
&=&\Re \sum_{k=k^{\prime} + k^{\prime\prime}} n_0  c_s^2 \langle ik_y \rho_s \phi_k^* u_k \rangle \nonumber \\
&=& n_0 \left( - \chi^{QL}\frac{\partial \langle U_{\|} \rangle }{\partial r} + \Pi_{r,\|}^{QL,res} \right),
\end{eqnarray}
with quasilinear diffusivity
\begin{equation}
\chi^{QL} = \rho_s c_s \displaystyle \sum_{k=k^{\prime} + k^{\prime\prime}} h_k I_k \nonumber\\,
\end{equation}
and quasilinear residual stress
\begin{equation}
\Pi_{r,\|}^{QL,res} = \displaystyle - c_s^2 \sum_{k=k^{\prime} + k^{\prime\prime}} h_k \frac{\Delta^2}{L_s L_I} I_k.\nonumber\\
\end{equation}
Here, $ \displaystyle h_k =\frac{k_y^2\rho_sc_s \left| \gamma_{k,NL2} \right|}{\left(\omega_k^2 + \gamma_{k,NL2}^2\right)} $. Parallel symmetry breaking induced by fluctuation intensity gradient \cite{Gurcan2006}  is used for the quasilinear residual stress. The quasilinear diffusivity is positive definite. Different from the nonlinear residual stress, the sign of quasilinear residual stress depends on both $ L_I $ and $ L_s $. $ \Pi_{r,\|}^{QL,res} $ is negative for increasing intensity profile and normal magnetic shear.

Before comparing the nonlinear results with the quasilinear ones, we clarify the orderings of typical parameters that we will take in the following. The relative fluctuation amplitude from mixing length estimate, i.e., $\phi_k \sim \frac{1}{k_{\perp} L_n} $ is used. For the spatial scales, $k_x \sim k_y \sim k_{\perp} \sim 1/\triangle$, $ L_n \sim L_I \sim L$, and $\displaystyle \left( \frac{\rho_s}{\Delta} \right)^2 \sim \left( \frac{\Delta}{L} \right)^2  \sim \frac{L}{L_s} \sim \epsilon$ are used, with $\epsilon\ll1$ a small ordering parameter, which are consistent with TORPEX parameters.\cite{Labit11} For the temporal scale, normalized real frequency is order of $\frac{\omega_k}{(k_y c_s)} \sim \frac{\rho_s}{L} \sim \epsilon$. The triad interaction time can be estimated as the inverse of nonlinear damping rate, because the frequency mismatch is much smaller than the nonlinear damping rate, as mentioned before. It was shown that the order of magnitude of the nonlinear damping rate for vorticity equation could be estimated as $\gamma_{k,1}^{NL}\sim \displaystyle \frac{k_{\perp}^3 \rho_s^3}{1+k_{\perp}^2\rho_s^2} k_y c_s \phi_k \sim k_{\perp}^2 \rho_s^2 \omega_k$.\cite{Diamond book} Comparing the nonlinear terms in vorticity equations and the parallel momentum equation, one can divide $\gamma_{k,1}^{NL}$  by a factor of $k_{\perp}^2 \rho_s^2$ to estimate $\gamma_{k,2}^{NL}$, i.e., $\gamma_{k,2}^{NL}\sim \omega_k$. The mean flow is one order smaller than the ion acoustic velocity, i.e., $\langle U_{\|}\rangle /c_s \sim \epsilon$. Then we can estimate the order of magnitude of the nonlinear and quasilinear diffusivity and residual stress based on above orderings. The results are listed in Table~\ref{comparison}.
\begin{table}\caption{Comparison of nonlinear and quasilinear results. \\
}\label{comparison}
\begin{tabular}[b]{ccc}
  \hline \hline &\qquad\qquad\qquad\qquad diffusivity($\rho_sc_s$)  \quad \quad  &\qquad\qquad\qquad\qquad residual stress ($c_s^2$) \quad \\
\hline  $\langle \tilde{v}_r \tilde{u}_{\|} \rangle$ &\qquad\qquad\qquad\qquad $\epsilon^{1/2}$ \qquad\qquad &\qquad\qquad\qquad\qquad $-\epsilon^{5/2}$ \quad\\
 $\langle \tilde{v}_r^{(c)} \tilde{n} \tilde{u}_{\|} \rangle + \langle \tilde{v}_r  \tilde{n}^{(c)} \tilde{u}_{\|} \rangle $ &\qquad\qquad\qquad $\epsilon^{3/2}$ \quad &\qquad\qquad\qquad\qquad $-\epsilon^{7/2}$ \quad \\
$ \langle \tilde{v}_r \tilde{n} \tilde{u}_{\|}^{(c)} \rangle $ & \qquad\qquad\qquad $\epsilon^{5/2}$ \quad &\qquad\qquad\qquad\qquad $\epsilon^{5/2}$ \quad \\
 \hline \hline
\end{tabular}
\end{table}

From Table~\ref{comparison}, we can see that the nonlinear diffusivity is smaller than the quasilinear one. However, the leading order nonlinear residual stress is in the same order as quasi-linear residual stress, but with opposite sign for increasing fluctuation intensity profile. Note that the dominant contribution to the nonlinear residual stress comes from $ \langle \tilde{v}_r \tilde{n} \tilde{u}_{\|}^{(c)} \rangle $     which is due to the coherent component of $\tilde {u}_{\|}$. This is because the order of the nonlinear interaction coefficient for $\tilde {u}_{\|}$ is higher than that for $\phi_k$ by an order $ k_{\perp}^2 \rho_s^2 \sim \frac{\rho_s^2}{\triangle^2} \sim \epsilon $.  Moreover, the term proportional to $\frac{\partial}{\partial r}u_{k\prime}$ or $\frac{\partial}{\partial r}u_{k\prime\prime}$ in $M^{2}_{k,k^{\prime},k^{\prime\prime}}$ results in the leading order nonlinear residual stress. The details of calculation can be found in Appendix B. The radial derivative of $ u_k $ contains the radial derivative of $ k_{\parallel} $, due to the radial position dependence of $ k_{\parallel} $. We also note that an asymmetric parallel fluctuation spectrum is not necessary for non-zero, nonlinear residual stress due to the radial derivative of $ k_{\parallel} $. This is in contrast to the quasilinear residual stress, in which an asymmetric parallel fluctuation spectrum is required. For instance, the factor $\frac{\triangle^2}{L_I L_s}$ in the quasilinear residual stress is induced by fluctuation intensity gradient symmetry breaking, which is order of $\epsilon^{2}$. This is a partial reason why nonlinear residual stress can be comparable to the quasilinear one. Another reason is the relative fluctuation amplitude is large in strong turbulence, i.e., $|\phi_k|^2 \sim \frac{\triangle^2}{L^2} \sim \epsilon$. This is different from weak turbulence, for which $|\phi_k|^2 \sim \epsilon^2$.

\section{Summary and discussion}

In the present work, we have derived the nonlinear parallel momentum flux for a three dimensional coupled drift waves and ion acoustic waves system. A Markovian approximation has been used for closure modelling. We estimate the triad interaction times for strong turbulence conditions. We also compared the nonlinear results with the quasilinear ones. It is shown that the nonlinear diffusivity is smaller than the quasilinear one. The leading order nonlinear residual stress can be comparable to quasilinear one based on the orderings we choose. This indicates that  taking into account nonlinear wave-wave coupling effects on parallel intrinsic rotation is important. It is known that non-zero quasilinear residual stress requires symmetry breaking, such as fluctuation intensity gradient\cite{Gurcan2006} which we adopt in this work. However, in contrast to quasilinear theory, we find that an asymmetric parallel fluctuation spectrum is not required for a non-zero nonlinear residual stress.

According to our theoretical results, the nonlinear residual stress is not negligible for strong turbulence. It supports the strong nonlinear parallel momentum flux induced by blob ejection which was observed in TORPEX expermiment.\cite{Labit11} In general, nonlinear momentum flux is also of potential relevance to tokamak edge region where similar blobs in L-mode are also observed. Therefore, it may be needed to include the effects of nonlinear residual stress induced by blob ejection on intrinsic rotation in tokamak experiments. Recent experiments on ASDEX-U found the triple fluctuation term, $ \tilde{v}_r \tilde{n} \tilde{v}_{pol} $ is dominant as compared to the poloidal Reynolds stress in turbulent poloidal momentum flux induced by ELM bursts during an H-mode discharge.\cite{Muller 2011} In our future work, we will extend this work to focus on theoretical models of the nonlinear turbulent poloidal momentum transport. We also plan to compare nonlinear residual stress models with J-TEXT measurements of intrinsic torque and blob populations.

Finally, we note that both resonant particle momentum diffusivity and residual stress can be calculated systematically to $O(\phi_k^4)$ in perturbation theory, for weak turbulence. The analysis follows Manheimer and Dupree\cite{ManheimerandDupree68} and is similar to that for anomalous heating.\cite{Zhao2012} Though nominally higher order in perturbation theory, the results are \emph{not} negligible, on account of differing resonant particle populations at different phase speeds.



\section*{Acknowledgments}

We are grateful to J. Q. Dong, J. Cheng, and Z.P. Chen for useful discussions. This work was supported by the MOST of China under Contract No.~2013GB112002, the NSFC Grant No.~11305071, and U.S. DOE Contract No. DE-FG02-04ER54738.

\appendix
\section{Linear response function of three dimensional Hasegawa-Mima system}
The linearization of three dimensional Hasegawa-Mima system can be proceeded as following.
\begin{equation}
H=\left[
  \begin{array}{cc}
   \displaystyle  i \frac{\omega_{*n}}{1+k_{\perp}^2{\rho_s^2}} & \displaystyle \frac{ik_{\|}c_s}{1+k_{\perp}^2{\rho_s^2}} \\
  \displaystyle ik_{\|}c_s  & 0 \\
  \end{array}
\right] \equiv \left[
  \begin{array}{cc}
   a & b \\
  c & 0 \\
  \end{array}
\right] ,\nonumber
\end{equation}
where, we have neglected  $ ik_y \rho_s \frac{\partial}{\partial r} \langle U_{\|} \rangle $.

\begin{equation}
\left|
\begin{array}{cc}
a-\lambda & b  \\
c & -\lambda  \\

\end{array}
\right| = 0 \Rightarrow  \lambda^2-a\lambda-bc=0 \ . \nonumber
\end{equation}

\begin{equation}
\lambda= \frac{\displaystyle i \frac{\omega_{*n}}{1+k_{\perp}^2{\rho_s^2}}\pm \sqrt{\left( i \frac{\omega_{*n}}{1+k_{\perp}^2{\rho_s^2}} \right)^2+ 4 \frac{ik_{\|}c_s}{1+k_{\perp}^2{\rho_s^2}} ik_{\|}c_s    }}{2} \ .\nonumber
\end{equation}


\begin{equation}
\lambda_1 \cong i \frac{\omega_{*n}}{1+k_{\perp}^2{\rho_s^2}}+i\frac{k_{\|}^2 c_s^2}{\omega_{*n}} , \lambda_2 \cong -i\frac{k_{\|}^2 c_s^2}{\omega_{*n}}.
\end{equation}
The transformation matrixes are
\begin{equation}
P = \left[
  \begin{array}{cc}
  b & b \\
   \lambda_1-a &  \lambda_2-a \\
  \end{array}
\right], \text{and} \ P^{-1} =\frac{1}{b(\lambda_2-\lambda_1)} \left[
  \begin{array}{cc}
 \lambda_2-a & -b \\
   a-\lambda_1 & b  \\
  \end{array}
\right].  \nonumber
\end{equation}

By direct diagonalization of matrix $H$ in Eq.~(\ref{etak}) and with the help of Eq.~(\ref{Markovian approximation}), we can get the equation,
\begin{equation}
\frac{\partial}{\partial t} \eta_k^{\prime \alpha} + \left(\lambda_{k,\alpha} + \gamma_{k.\alpha}^{NL} \right) \eta_k^{\prime \alpha} = 2 M_{k,k^{\prime},k^{\prime\prime}}^{\prime \alpha}+ F_{k, \alpha}^{\prime}.
\end{equation}
Then, the above equation can be solved ss
\begin{equation}\label{etakprime coherent}
\eta_k^{\prime \alpha (c)}=\int_{-\infty} ^t dt^{\prime}exp \left[  \left(\lambda_{k,\alpha} + \gamma_{k.\alpha}^{NL} \right) \left(t^{\prime}-t \right) \right] \left( 2 M_{k,k^{\prime},k^{\prime\prime}}^{\prime \alpha} + F_{k, \alpha}^{\prime}  \right) ,
\end{equation}
where $\eta_k^{\prime \alpha}=P_{\alpha, \beta}^{-1}\eta_k^{\beta},      M_{k,k^{\prime},k^{\prime\prime}}^{\prime \alpha}=P_{\alpha, \beta}^{-1}M_{k,k^{\prime},k^{\prime\prime}}^\beta , F_{k, \alpha}^{\prime}=P_{\alpha, \beta}^{-1}F_{k, \beta}. \nonumber $

Because $F_{k,\alpha}$ is fast fluctuating force, which does not contribute to coherent parts of the beat mode, we can neglect it here. Multiplication of Eq.~(\ref{etakprime coherent}) by matrix $P$ on the left, we can get the coherent component of the beat mode
\begin{equation}
\eta_k^{\alpha (c)}= P_{\alpha,\beta}\eta_k^{\prime \beta (c)} .
\end{equation}
The explicit expressions can be written as
\begin{align}\label{phi coherent}
\phi_k^{(c)} &= \int_{-\infty} ^t dt^{\prime}exp \left[ \left( i\omega_{k 1} + \gamma_{k 1}^{NL} \right) (t^{\prime}-t) \right]P_{1 1} \left[ P_{11 }^{-1} M_{k,k^{\prime},k^{\prime\prime}}^1(t^{\prime})+ P_{12 }^{-1} M_{k,k^{\prime},k^{\prime\prime}}^2(t^{\prime}) \right] \nonumber \\
&+\int_{-\infty} ^t dt^{\prime}exp \left[ \left( i\omega_{k 2} + \gamma_{k 2}^{NL} \right) (t^{\prime}-t) \right]P_{1 2} \left[ P_{21 }^{-1} M_{k,k^{\prime},k^{\prime\prime}}^1(t^{\prime})+ P_{22 }^{-1} M_{k,k^{\prime},k^{\prime\prime}}^2(t^{\prime}) \right], \nonumber
\end{align}
\begin{eqnarray}\label{u coherent}
 u_k^{(c)} &= \int_{-\infty} ^t dt^{\prime}exp \left[ \left( i\omega_{k 1} + \gamma_{k 1}^{NL} \right) (t^{\prime}-t) \right]P_{2 1} \left[ P_{11 }^{-1} M_{k,k^{\prime},k^{\prime\prime}}^1(t^{\prime})+ P_{12 }^{-1} M_{k,k^{\prime},k^{\prime\prime}}^2(t^{\prime}) \right] \nonumber \\
&+\int_{-\infty} ^t dt^{\prime}exp \left[ \left( i\omega_{k 2} + \gamma_{k 2}^{NL} \right) (t^{\prime}-t) \right]P_{2 2} \left[ P_{21 }^{-1} M_{k,k^{\prime},k^{\prime\prime}}^1(t^{\prime})+ P_{22 }^{-1} M_{k,k^{\prime},k^{\prime\prime}}^2(t^{\prime}) \right]. \nonumber
\end{eqnarray}
Then, we can rewrite them compactly in the form of matrix as Eqs.~(\ref{coherent})and (\ref{response function}) easily. The matrix in response functions are $r_k^{\alpha\beta1}=P_{\alpha1}P_{1\beta}^{-1}$, and $r_k^{\alpha\beta2}=P_{\alpha2}P_{2\beta}^{-1}$ respectively.

\section{Calculation of nonlinear momentum flux}
More detailed calculation of nonlinear contribution is given this section.
The linearization of Eq.~(\ref{uk}) can be written as
\begin{equation}
u_k=\frac{\displaystyle k_{\parallel}c_s-k_y  \rho_s \frac{\partial \langle U_{\|} \rangle}{\partial r} }{\omega_k+i\gamma_{k,NL2}}\phi_k .
\end{equation}
Because $\omega_{k2}$ resulted from correction of the parallel compression is much smaller than $\omega_{k1}$, we only take $\omega_{k1}$ here, and omit the subscript 1 for simplicity in the following calculation.
First, we present the result of $\Pi_{r,\| 1}^{NL}$ without tedious calculations.
\begin{eqnarray}
\Pi_{r,\| 1}^{NL} &=& \langle \tilde{v}_r^{(c)} \tilde{n} \tilde{u}_{\|} \rangle + \langle \tilde{v}_r  \tilde{n}^{(c)} \tilde{u}_{\|} \rangle \nonumber \\
&=& n_0 c_s^2 \Re\sum_{k=k^{\prime} + k^{\prime\prime}}\left[\left(ik_y-ik_y^{\prime} \right)\rho_s \int_{-\infty}^{t}dt^{\prime} \langle R_k^{1\beta}(t,t^{\prime}) M_{k,k^{\prime},k^{\prime\prime}}^{\beta*}(t^{\prime}) n_{k^{\prime}}(t) u_{k^{\prime\prime}}(t)\rangle \right] \nonumber\\
&=& n_0 c_s^2 \Re\sum_{k=k^{\prime} + k^{\prime\prime}}ik_y^{\prime\prime} \rho_s \int_{-\infty}^{t}dt^{\prime} exp \left[ \left(-i\omega_{k1} + \gamma_{k 1}^{NL} \right) (t^{\prime}-t) \right] \langle M_{k,k^{\prime},k^{\prime\prime}}^{1*}(t^{\prime}) \phi_{k^{\prime}}(t) u_{k^{\prime\prime}}(t)\rangle \nonumber \\
&+& n_0 c_s^2 \Re\sum_{k=k^{\prime} + k^{\prime\prime}} ik_y^{\prime\prime} \rho_s \int_{-\infty}^{t}dt^{\prime} exp \left[ \left(-i\omega_{k1} + \gamma_{k 1}^{NL} \right) (t^{\prime}-t) \right] \langle \frac{k_{\parallel}c_s}{\omega_{*n}} M_{k,k^{\prime},k^{\prime\prime}}^{2*}(t^{\prime}) \phi_{k^{\prime}}(t) u_{k^{\prime\prime}}(t)\rangle \nonumber \\
&=&\frac{1}{4}n_0 \rho_s c_s\sum_{k=k^{\prime} + k^{\prime\prime}} I_{k\prime}I_{k\prime\prime} \frac{\partial \left\langle U_{\parallel} \right\rangle}{\partial{r}} \left\lbrace -\frac{\tau_{c1}\omega_{ci}}{(1+k_{\perp}^2 \rho_s^2)}g_{k\prime\prime}\frac{\rho_s}{L_I} A_{k\prime,k\prime\prime} \right.\nonumber \\
&&\left. +\frac{L_n}{L_s}  \tau_{c2}\omega_{ci} \left[ 2k_y^{\prime\prime2}\rho_s^2  \frac{p_{k\prime\prime}\Delta^{\prime\prime2}}{L_sL_I}+\frac{\Delta^{\prime2}+\Delta^{\prime\prime2}}{L_sL_I}\Bigg (h_{k\prime}h_{k\prime\prime} -\frac{3}{2} k_y^{\prime\prime2}\rho_s^2 g_{k\prime}g_{k\prime\prime} \Bigg)\right]\right\rbrace \nonumber \\
&-&\frac{1}{4}n_0 c_s^2 \sum_{k=k^{\prime}+k^{\prime\prime}} I_{k\prime}I_{k\prime\prime} \left[ \frac{\tau_{c1}\omega_{ci}}{(1+k_{\perp}^2 \rho_s^2)}  g_{k\prime\prime}A_{k\prime,k\prime\prime}\frac{\Delta^{\prime\prime2}}{L_sL_I} \frac{\rho_s}{L_I} + 2\frac{L_n}{L_s}\tau_{c2}\omega_{ci} k_y^{\prime\prime2}\rho_s^2g_{k\prime}g_{k\prime\prime}\frac{\Delta^{\prime\prime2}}{L_s^2}  \right.\nonumber \\
&&\left.- \tau_{c2}\omega_{ci} \frac{\Delta^{\prime2}\Delta^{\prime\prime2}}{L_s^2L_I^2} \frac{L_n}{L_s} \Bigg (h_{k\prime}h_{k\prime\prime} +k_y^{\prime\prime2}\rho_s^2 p_{k\prime\prime}-3k_y^{\prime\prime2} \rho_s^2 g_{k\prime}g_{k\prime\prime} \Bigg) \right] \nonumber \\
&=&- n_0 \chi_1^{NL}\frac{\partial \langle U_{\|} \rangle }{\partial r} + n_0 \Pi_{r,\| 1}^{NL,res},
\end{eqnarray}
with
\begin{eqnarray}
\chi_1^{NL}&=&\frac{1}{4} \rho_s c_s\sum_{k=k^{\prime} + k^{\prime\prime}} I_{k\prime}I_{k\prime\prime} \left\lbrace \frac{\tau_{c1}\omega_{ci}}{(1+k_{\perp}^2 \rho_s^2)} \frac{\rho_s}{L_I} g_{k\prime\prime} A_{k\prime,k\prime\prime} \right.\nonumber \\
&&\left. - \tau_{c2}\omega_{ci} \frac{L_n}{L_s} \left[ 2k_y^{\prime\prime2}\rho_s^2 p_{k\prime\prime} \frac{\Delta^{\prime\prime2}}{L_sL_I}+\frac{\Delta^{\prime2}+\Delta^{\prime\prime2}}{L_sL_I}\Bigg (h_{k\prime}h_{k\prime\prime} -\frac{3}{2} k_y^{\prime\prime2}\rho_s^2 g_{k\prime}g_{k\prime\prime} \Bigg)\right]\right\rbrace, \nonumber
\end{eqnarray}
and
\begin{eqnarray}
\Pi_{r,\| 1}^{NL,res}&=&-\frac{1}{4} c_s^2 \sum_{k=k^{\prime}+k^{\prime\prime}} I_{k\prime}I_{k\prime\prime} \left[ \frac{\tau_{c1}\omega_{ci}}{(1+k_{\perp}^2 \rho_s^2)}  g_{k\prime\prime}A_{k\prime,k\prime\prime}\frac{\Delta^{\prime\prime2}}{L_sL_I} \frac{\rho_s}{L_I} + 2\frac{L_n}{L_s}\tau_{c2}\omega_{ci} k_y^{\prime\prime2}\rho_s^2g_{k\prime}g_{k\prime\prime}\frac{\Delta^{\prime\prime2}}{L_s^2}  \right.\nonumber \\
&&\left.- \tau_{c2}\omega_{ci} \frac{\Delta^{\prime2}\Delta^{\prime\prime2}}{L_s^2L_I^2} \frac{L_n}{L_s} \Bigg (h_{k\prime}h_{k\prime\prime} +k_y^{\prime\prime2}\rho_s^2 p_{k\prime\prime}-3k_y^{\prime\prime2} \rho_s^2 g_{k\prime}g_{k\prime\prime} \Bigg) \right]. \nonumber
\end{eqnarray}
Here, the dimensionless parameters are $A_{k^{\prime},k^{\prime\prime}} \equiv k_y^{\prime\prime2} \left(k_{\perp}^{\prime2} - k_{\perp}^{\prime\prime2} + 2k_{x}^{\prime2}\right)\rho_s^4 $, $\displaystyle g_k \equiv \frac{k_yc_s \omega_k}{\left(\omega_k^2 + \gamma_{k,NL2}^2\right)} $, $\displaystyle h_k \equiv \frac{k_y^2\rho_sc_s \gamma_{k,NL2}}{\left(\omega_k^2 + \gamma_{k,NL2}^2\right)} $ and $ \displaystyle p_k \equiv \frac{k_y^2c_s^2 }{\left(\omega_k^2 + \gamma_{k,NL2}^2\right)} $. The first lines in $\chi_1^{NL}$ and $\Pi_{r,\| 1}^{NL,res}$ are leading order.

The calculation of $\Pi_{r,\| 2}^{NL}$ is written as follows.
\begin{eqnarray} \label{tri2}
\Pi_{r,\| 2}^{NL} &=& \langle \tilde{v}_r \tilde{n} \tilde{u}_{\|}^{(c)} \rangle \nonumber \\
&=& n_0 c_s^2\ \Re\sum_{k=k^{\prime} + k^{\prime\prime}}-ik_y^{\prime} \rho_s \int_{-\infty}^{t}dt^{\prime} \langle R_k^{2 \beta*}(t,t^{\prime}) M_{k,k^{\prime},k^{\prime\prime}}^{\beta*}(t^{\prime})  \phi_{k^{\prime}}(t) n_{k^{\prime\prime}}(t)\rangle \nonumber\\
&=& n_0 c_s^2 \Re\sum_{k=k^{\prime} + k^{\prime\prime}}-ik_y^{\prime} \rho_s \int_{-\infty}^{t}dt^{\prime} exp \left[ \left( -i\omega_{k2} + \gamma_{k 2}^{NL} \right) (t^{\prime}-t) \right] \langle  M_{k,k^{\prime},k^{\prime\prime}}^{2*}(t^{\prime}) \phi_{k^{\prime}}(t) n_{k^{\prime\prime}}(t)\rangle \nonumber \\
&+&  n_0 c_s^2 \Re\sum_{k=k^{\prime} + k^{\prime\prime}}-ik_y^{\prime} \rho_s \int_{-\infty}^{t}dt^{\prime} exp \left[ \left( -i\omega_{k1} + \gamma_{k 1}^{NL} \right) (t^{\prime}-t) \right] \langle \frac{k_{\parallel}c_s(1+k_{\perp}^2 \rho_s^2)}{\omega_{*n}} M_{k,k^{\prime},k^{\prime\prime}}^{1*}(t^{\prime}) \phi_{k^{\prime}}(t) n_{k^{\prime\prime}}(t)\rangle \nonumber \\
&+& n_0 c_s^2 \Re\sum_{k=k^{\prime} + k^{\prime\prime}}-ik_y^{\prime} \rho_s \int_{-\infty}^{t}dt^{\prime} exp \left[ \left( -i\omega_{k2} + \gamma_{k 2}^{NL} \right) (t^{\prime}-t) \right] \langle \frac{k_{\parallel}^2c_s^2(1+k_{\perp}^2 \rho_s^2)}{\omega_{*n}^2} M_{k,k^{\prime},k^{\prime\prime}}^{2*}(t^{\prime}) \phi_{k^{\prime}}(t) n_{k^{\prime\prime}}(t)\rangle \nonumber \\
&=&-\frac{1}{2}n_0 c_s^2 \Re\sum_{k=k^{\prime} + k^{\prime\prime}} \tau_{c2}\omega_{ci}k_y^{\prime 2} \rho_s^2 \left[\langle \phi_{k\prime}^{*}(t)\phi_{k\prime}(t) \rangle \langle \rho_s \frac{\partial u_{k \prime\prime}^{*}(t)}{\partial{r}} n_{k \prime\prime}(t) \rangle - \langle u_{k\prime}^{*}(t)\phi_{k\prime}(t) \rangle \langle \rho_s \frac{\partial \phi_{k \prime\prime}^{*}(t)}{\partial{r}} n_{k \prime\prime}(t) \rangle \right] \nonumber \\
&+&\frac{1}{2}n_0 c_s^2 \Re\sum_{k=k^{\prime} + k^{\prime\prime}} \tau_{c1}\omega_{ci} \frac{k_{\parallel}c_s(1+k_{\perp}^2 \rho_s^2)}{\omega_{*n}} k_y^{\prime 2} \left( k_{\perp}^{\prime\prime 2}-k_{\perp}^{\prime\ 2} + 2k_x^{\prime\prime 2} \right)\rho_s^4  \langle \phi_{k\prime}^{*}(t)\phi_{k\prime}(t) \rangle \langle \rho_s \frac{\partial \phi_{k \prime\prime}^{*}(t)}{\partial{r}} n_{k \prime\prime}(t) \rangle \nonumber \\
&-&\frac{1}{2}n_0 c_s^2 \Re\sum_{k=k^{\prime} + k^{\prime\prime}} \tau_{c2}\omega_{ci} \frac{k_{\parallel}^2c_s^2(1+k_{\perp}^2 \rho_s^2)}{\omega_{*n}^2} k_y^{\prime 2} \rho_s^2 \left[\langle \phi_{k\prime}^{*}(t)\phi_{k\prime}(t) \rangle \langle \rho_s \frac{\partial u_{k \prime\prime}^{*}(t)}{\partial{r}} n_{k \prime\prime}(t) \rangle \right. \nonumber \\
&&-\left. \langle u_{k\prime}^{*}(t)\phi_{k\prime}(t) \rangle \langle \rho_s \frac{\partial \phi_{k \prime\prime}^{*}(t)}{\partial{r}} n_{k \prime\prime}(t) \rangle \right].
\end{eqnarray}
As we can see, we have written $\Pi_{r,\| 2}^{NL}$ into three components, and we will calculate each of them one by one as following.
\begin{eqnarray} \label{firstcomponent1}
&&\sum_{k=k^{\prime} + k^{\prime\prime}} \langle \rho_s \frac{\partial u_{k \prime\prime}^{*}(t)}{\partial{r}} n_{k \prime\prime}(t) \rangle \nonumber \\
&=& \sum_{k=k^{\prime} + k^{\prime\prime}}\frac{c_s}{\rho_s} \left[ \frac{\displaystyle (\frac{x^{\prime\prime}}{L_s}-\frac{\rho_s}{c_s} \frac{\partial \left\langle U_{\parallel} \right\rangle}{\partial{r}} )k_y^{\prime\prime}\rho_s}{\omega_{k\prime\prime}-i\gamma_{k\prime\prime2}^{NL}}\left\langle \rho_s \frac{\partial \phi_{k\prime\prime}^{*}}{\partial{r}}\phi_{k\prime\prime} \right\rangle - \frac{\rho_s \displaystyle \frac{1}{L_s}k_y^{\prime\prime}\rho_s}{\omega_{k\prime\prime}-i\gamma_{k\prime\prime2}^{NL}}\left\langle \phi_{k\prime\prime}^{*}\phi_{k\prime\prime} \right\rangle \right] \nonumber \\
&=&\sum_{k=k^{\prime} + k^{\prime\prime}}-\frac{1}{2}\frac{ \displaystyle k_y^{\prime\prime}\rho_s I_{k\prime\prime}}{\omega_{k\prime\prime}-i\gamma_{k\prime\prime2}^{NL}}\left( 2 \frac{c_s}{L_s}+\frac{\rho_s}{L_I} \frac{\partial \left\langle U_{\parallel} \right\rangle}{\partial{r}} \right).
\end{eqnarray}
Here, $ k_{\parallel}= k_y \displaystyle \frac{x}{L_s} $ is used. $L_s$ is the magnetic shear scale length, $x=r_0-r$, where $r_0$ is the radial location of resonant surface. So, the second term in the second line comes from $\displaystyle \frac{\partial}{\partial{r}}k_{\parallel}=-\frac{k_y}{L_s}$. Note that parallel asymmetric fluctuation spectrum is not required for the second term.
\begin{eqnarray}\label{firstcomponent2}
\sum_{k=k^{\prime} + k^{\prime\prime}} \langle u_{k\prime}^*(t) \phi_{k \prime}(t) \rangle &=& \sum_{k=k^{\prime} + k^{\prime\prime}} \left\langle \frac{c_s}{\rho_s} \frac{\displaystyle (\frac{x^{\prime}}{L_s}-\frac{\rho_s}{c_s} \frac{\partial \left\langle U_{\parallel} \right\rangle}{\partial{r}} )k_y^{\prime}\rho_s}{\omega_{k\prime}-i\gamma_{k\prime2}^{NL}} \phi_{k\prime}^{*}(t)\phi_{k\prime(t)}          \right\rangle \nonumber \\
&=&-\sum_{k=k^{\prime} + k^{\prime\prime}}\frac{ \displaystyle k_y^{\prime}\rho_s I_{k\prime}}{\omega_{k\prime}-i\gamma_{k\prime2}^{NL}}\left( \frac{c_s}{\rho_s}\frac{\Delta^{\prime2}}{L_sL_I}+ \frac{\partial \left\langle U_{\parallel} \right\rangle}{\partial{r}} \right).
\end{eqnarray}
According to Eqs.~(\ref{firstcomponent1}) and (\ref{firstcomponent2}) we can get the first component of $\Pi_{r,\| 2}^{NL}$, i.e., Eq.~(\ref{tri2}),
\begin{eqnarray}
&&-\frac{1}{2}n_0 c_s^2 \Re\sum_{k=k^{\prime} + k^{\prime\prime}} \tau_{c2}\omega_{ci}k_y^{\prime 2} \rho_s^2 \left[\langle \phi_{k\prime}^{*}(t)\phi_{k\prime}(t) \rangle \langle \rho_s \frac{\partial u_{k \prime\prime}^{*}(t)}{\partial{r}} n_{k \prime\prime}(t) \rangle - \langle u_{k\prime}^{*}(t)\phi_{k\prime}(t) \rangle \langle \rho_s \frac{\partial \phi_{k \prime\prime}^{*}(t)}{\partial{r}} n_{k \prime\prime}(t) \rangle \right] \nonumber \\
&=&-\frac{1}{2}n_0 c_s^2 \Re\sum_{k=k^{\prime} + k^{\prime\prime}} \tau_{c2}\omega_{ci}k_y^{\prime 2} \rho_s^2 \left[ -\frac{1}{2}I_{k\prime}\frac{ \displaystyle k_y^{\prime\prime}\rho_s I_{k\prime\prime}}{\omega_{k\prime\prime}-i\gamma_{k\prime\prime2}^{NL}}\left( 2 \frac{c_s}{L_s}+\frac{\rho_s}{L_I} \frac{\partial \left\langle U_{\parallel} \right\rangle}{\partial{r}} \right) \right. \nonumber \\
&&+\left. \frac{ \displaystyle k_y^{\prime}\rho_s I_{k\prime}}{\omega_{k\prime}-i\gamma_{k\prime2}^{NL}}\left( \frac{c_s}{\rho_s}\frac{\Delta^{\prime2}}{L_sL_I}+ \frac{\partial \left\langle U_{\parallel} \right\rangle}{\partial{r}} \right)\frac{\rho_sI_{k\prime\prime}}{2L_I} \right] \nonumber \\
&=&-\frac{1}{4}n_0  c_s^2 \sum_{k=k^{\prime} + k^{\prime\prime}} \tau_{c2}\omega_{ci}I_{k\prime}I_{k\prime\prime} \left[\frac{\rho_s}{L_I} \left(g_{k\prime}-g_{k\prime\prime}\right) \frac{\rho_s}{c_s} k_y^{\prime2} \rho_s^2 \frac{\partial \left\langle U_{\parallel} \right\rangle}{\partial{r}} +\frac{\rho_s}{L_I} k_y^{\prime2}g_{k\prime}\frac{\Delta^{\prime2}}{L_sL_I} - 2 \frac{\rho_s}{L_s}k_y^{\prime2}g_{k\prime\prime}\right]. \nonumber
\end{eqnarray}
Now, we turn to calculation of the second component of $\Pi_{r,\| 2}^{NL}$.
\begin{eqnarray}
&&\frac{1}{2}n_0 c_s^2 \Re\sum_{k=k^{\prime} + k^{\prime\prime}} \tau_{c1}\omega_{ci} \frac{k_{\parallel}c_s(1+k_{\perp}^2 \rho_s^2)}{\omega_{*n}} k_y^{\prime 2} \left( k_{\perp}^{\prime\prime 2}-k_{\perp}^{\prime\ 2} + 2k_x^{\prime\prime 2} \right)\rho_s^4  \langle \phi_{k\prime}^{*}(t)\phi_{k\prime}(t) \rangle \langle \rho_s \frac{\partial \phi_{k \prime\prime}^{*}(t)}{\partial{r}} n_{k \prime\prime}(t) \rangle \nonumber \\
&=&\frac{1}{2}n_0 c_s^2 \Re\sum_{k=k^{\prime} + k^{\prime\prime}} \tau_{c1}\omega_{ci} \frac{L_n \left( x^{\prime} +x^{\prime\prime} \right) (1+k_{\perp}^2 \rho_s^2)}{L_s\rho_s} A_{k\prime\prime,k\prime} \langle \phi_{k\prime}^{*}(t)\phi_{k\prime}(t) \rangle \langle \rho_s \frac{\partial \phi_{k \prime\prime}^{*}(t)}{\partial{r}} n_{k \prime\prime}(t) \rangle \nonumber \\
&=&-\frac{1}{2}n_0 c_s^2 \sum_{k=k^{\prime} + k^{\prime\prime}} \tau_{c1}\omega_{ci} \frac{L_n (1+k_{\perp}^2 \rho_s^2)}{L_s}  A_{k\prime\prime,k\prime} \frac{\Delta^{\prime2}}{2L_I^2}I_{k\prime} I_{k\prime\prime}, \nonumber
\end{eqnarray}
Noting that we have used the relation $\displaystyle\Re\sum_{k=k^{\prime} + k^{\prime\prime}} \langle x^{\prime\prime}  \rho_s \frac{\partial \phi_{k \prime\prime}^{*}(t)}{\partial{r}} n_{k \prime\prime}(t) \rangle=0$ in the calculation above.

By the same token, we can get the last component of $\Pi_{r,\| 2}^{NL}$ as follows:
\begin{eqnarray}
&&-\frac{1}{2}n_0 c_s^2 \Re\sum_{k=k^{\prime} + k^{\prime\prime}} \tau_{c2}\omega_{ci} \frac{k_{\parallel}^2c_s^2 (1+k_{\perp}^2 \rho_s^2) }{\omega_{*n}^2} k_y^{\prime 2} \rho_s^2 \left[ \langle \phi_{k\prime}^{*}(t)\phi_{k\prime}(t) \rangle \langle  \rho_s \frac{\partial u_{k \prime\prime}^{*}(t)}{\partial{r}} n_{k \prime\prime}(t) \rangle \right. \nonumber \\
&&\left.- \langle u_{k\prime}^{*}(t)\phi_{k\prime}(t) \rangle \langle \rho_s \frac{\partial \phi_{k \prime\prime}^{*}(t)}{\partial{r}} n_{k \prime\prime}(t) \rangle \right] \nonumber \\
&=&-\frac{1}{2}n_0 c_s^2 \Re\sum_{k=k^{\prime} + k^{\prime\prime}} \tau_{c2}\omega_{ci} \frac{L_n^2 (1+k_{\perp}^2 \rho_s^2) }{L_s^2\rho_s^2} (x^{\prime2}+x^{\prime\prime2}+2x^{\prime}x^{\prime\prime}) k_y^{\prime 2} \rho_s^2 \left[ \langle \phi_{k\prime}^{*}(t)\phi_{k\prime}(t) \rangle \langle \rho_s \frac{\partial u_{k \prime\prime}^{*}(t)}{\partial{r}} n_{k \prime\prime}(t) \rangle \right.\nonumber \\
&&\left.- \langle u_{k\prime}^{*}(t)\phi_{k\prime}(t) \rangle \langle \rho_s \frac{\partial \phi_{k \prime\prime}^{*}(t)}{\partial{r}} n_{k \prime\prime}(t) \rangle \right] \nonumber \\
&=& -\frac{1}{2}n_0 c_s^2 \Re\sum_{k=k^{\prime} + k^{\prime\prime}} \tau_{c2}\omega_{ci} I_{k\prime}I_{k\prime\prime} \frac{L_n^2}{L_s^2} (1+k_{\perp}^2 \rho_s^2) \Bigg \lbrace k_y^{\prime 2}(\Delta^{\prime2}+\Delta^{\prime\prime2}) \left[ -\frac{ \displaystyle k_y^{\prime\prime}\rho_s }{\omega_{k\prime\prime}-i\gamma_{k\prime\prime2}^{NL}}\left(  \frac{c_s}{L_s}+\frac{\rho_s}{2L_I} \frac{\partial \left\langle U_{\parallel} \right\rangle}{\partial{r}} \right) \right. \nonumber \\
 &&+\left. \frac{ \displaystyle k_y^{\prime}\rho_s }{\omega_{k\prime}-i\gamma_{k\prime2}^{NL}}\left( \frac{c_s}{\rho_s}\frac{\Delta^{\prime2}}{L_sL_I}+ \frac{\partial \left\langle U_{\parallel} \right\rangle}{\partial{r}} \right)  \frac{\rho_s}{2L_I} \right] -k_y^{\prime 2}\Delta^{\prime2} \frac{ \displaystyle k_y^{\prime\prime}\rho_s }{\omega_{k\prime\prime}-i\gamma_{k\prime\prime2}^{NL}}\frac{\Delta^{\prime\prime2}}{L_I^2} \frac{3c_s}{L_s} \Bigg\rbrace \nonumber \\
 &=&-\frac{1}{2}n_0 c_s^2\sum_{k=k^{\prime} + k^{\prime\prime}} \tau_{c2}\omega_{ci}I_{k\prime}I_{k\prime\prime} \frac{L_n^2 (1+k_{\perp}^2 \rho_s^2)}{L_s^2} \Bigg \lbrace k_y^{\prime 2} (\Delta^{\prime2}+\Delta^{\prime\prime2}) \left[ \left(g_k^{\prime}-g_k^{\prime\prime} \right) \frac{\rho_s}{c_s} \frac{\rho_s}{2L_I} \frac{\partial \left\langle U_{\parallel} \right\rangle}{\partial{r}} \right. \nonumber \\
&&+\left. g_k^{\prime} \frac{\rho_s}{2L_s} \frac{\Delta^{\prime2}}{L_I^2} -g_k^{\prime\prime}\frac{\rho_s}{L_s} \right] - g_k^{\prime\prime} \frac{3\rho_s}{L_s} k_y^{\prime2}\Delta^{\prime2} \frac{\Delta^{\prime\prime2}}{L_I^2} \Bigg \rbrace,
\end{eqnarray}
Here we have used the relation:
\begin{equation}
\displaystyle\Re\sum_{k=k^{\prime} + k^{\prime\prime}} \langle  x^{\prime\prime} \rho_s \frac{\partial u_{k \prime\prime}^{*}(t)}{\partial{r}} n_{k \prime\prime}(t) \rangle = g_k^{\prime\prime} I_{k\prime\prime} \frac{\Delta^{\prime\prime2}}{L_s} \frac{3\rho_s}{2L_I}. \nonumber
\end{equation}

Finally, we combine all the components of $\Pi_{r,\| 2}^{NL}$ to obtain
\begin{eqnarray} \label{Pi_2}
\Pi_{r,\| 2}^{NL} &=& -\frac{1}{4}n_0 \rho_s c_s  \sum_{k=k^{\prime} + k^{\prime\prime}} \tau_{c2}\omega_{ci}I_{k\prime}I_{k\prime\prime} \left(g_{k\prime}-g_{k\prime\prime}\right) \left[k_y^{\prime2} \rho_s^2 +k_y^{\prime2} (\Delta^{\prime2}+\Delta^{\prime\prime2}) \frac{L_n^2 (1+k_{\perp}^2 \rho_s^2)}{L_s^2} \right] \frac{\rho_s}{L_I} \frac{\partial \langle U_{\|} \rangle }{\partial r} \nonumber\\
&&+\frac{1}{2} n_0 c_s^2\sum_{k=k^{\prime} + k^{\prime\prime}} \tau_{c2}\omega_{ci}I_{k\prime}I_{k\prime\prime} \Bigg \lbrace g_{k\prime\prime} k_y^{\prime2} \rho_s^2 \frac{\rho_s}{L_s} - \frac{L_n^2 }{2L_s^2}\left(1+k_{\perp}^2 \rho_s^2 \right) \left[\frac{L_s(1+k_{\perp}^2 \rho_s^2)}{L_n k_{\perp}^2 \rho_s^2 } A_{k\prime\prime,k\prime} \frac{\Delta^{\prime2}}{L_I^2} \right.\nonumber \\
&& \left.- 2 g_{k\prime\prime} k_y^{\prime2} \left(\Delta^{\prime2}+\Delta^{\prime\prime2}\right) \frac{\rho_s}{L_s} + g_{k\prime} k_y^{\prime2} \rho_s^2 \frac{L_s^2}{L_n^2\left( 1+ k_{\perp}^2 \rho_s^2 \right)} \frac{\Delta^{\prime2}}{L_I^2} \frac{\rho_s}{L_s} \right] \Bigg \rbrace \nonumber \\
&=&- n_0 \chi_2^{NL} \frac{\partial \langle U_{\|} \rangle }{\partial r} + n_0 \Pi_{r,\| 2}^{NL,res} ,
\end{eqnarray}
with
\begin{eqnarray}
 \chi_2^{NL}&=& \frac{1}{4} \rho_s c_s \sum_{k=k^{\prime} + k^{\prime\prime}} \tau_{c2}\omega_{ci}I_{k\prime}I_{k\prime\prime}  \left(g_{k\prime}-g_{k\prime\prime}\right) \left[ k_y^{\prime2} \rho_s^2 +k_y^{\prime2} (\Delta^{\prime2}+\Delta^{\prime\prime2}) \frac{L_n^2 (1+k_{\perp}^2 \rho_s^2)}{L_s^2} \right]\frac{\rho_s}{L_I},  \nonumber
\end{eqnarray}
and
\begin{eqnarray}
 \Pi_{r,\| 2}^{NL,res}&=&\frac{1}{2} c_s^2\sum_{k=k^{\prime} + k^{\prime\prime}} \tau_{c2}\omega_{ci}I_{k\prime}I_{k\prime\prime} \Bigg \lbrace g_{k\prime\prime} k_y^{\prime2} \rho_s^2 \frac{\rho_s}{L_s} - \frac{L_n^2 }{2L_s^2}\left(1+k_{\perp}^2 \rho_s^2 \right) \left[\frac{L_s(1+k_{\perp}^2 \rho_s^2)}{L_n k_{\perp}^2 \rho_s^2 } A_{k\prime\prime,k\prime} \frac{\Delta^{\prime2}}{L_I^2} \right.\nonumber \\
&& \left.- 2 g_{k\prime\prime} k_y^{\prime2} \left(\Delta^{\prime2}+\Delta^{\prime\prime2}\right) \frac{\rho_s}{L_s} + g_{k\prime} k_y^{\prime2} \rho_s^2 \frac{L_s^2}{L_n^2\left( 1+ k_{\perp}^2 \rho_s^2 \right)} \frac{\Delta^{\prime2}}{L_I^2} \frac{\rho_s}{L_s} \right] \Bigg \rbrace. \nonumber
\end{eqnarray}
Note that the first terms of $\chi_2^{NL}$ and $\Pi_{r,\| 2}^{NL,res}$ are leading order. We have represented $ \tau_{c1} $ by $ \tau_{c2} $ by using the relation $\tau_{c2}= \displaystyle \frac{k_{\perp}^2 \rho_s^2}{(1+k_{\perp}^2 \rho_s^2)}\tau_{c1}$ which comes from the comparison between nonlinear terms in vorticity equations and those in parallel momentum equation.


\begin{thebibliography}{20}

\bibitem{A. Bondeson and D. J. Ward}
A. Bondeson and D. J. Ward, Phys. Rev. Lett. {\bf 72}, 2709 (1994).

\bibitem{R. Betti and J. P. Freidberg}
R. Betti and J. P. Freidberg, Phys. Rev. Lett. {\bf 74}, 2949 (1995).

\bibitem{S. M. Wolfe}
S. M. Wolfe, I. H. Hutchinson, R. S. Granetz, J. Rice, A. Hubbard, A. Lynn, P. Phillips, T. C. Hender, D. F. Howell, R. J. La Haye and J. T. Scoville, Phys. Plasmas {\bf 12}, 056110 (2005).

\bibitem{R. Fitzpatrick}
R. Fitzpatrick, Nucl. Fusion {\bf 33}, 1049 (1993).

\bibitem{E. J. Strait}
E. J. Strait, T. S. Taylor, A. D. Turnbull, J. R. Ferron, L. L. Lao, B. Rice, O. Sauter, S. J. Thompson, and D. Wroblewski, Phys. Rev. Lett. {\bf 74}, 2483 (1995).

\bibitem{B Goncalves}
B. Goncalves, C. Hidalgo, M. A. Pedrosa, C. Silva, R. Balbin, K. Erents, M. Hron, A. Loarte, G. Matthews, Plasma Phys. Control. Fusion {\bf 45}, 1627 (2003).

\bibitem{Rice08}
J. E. Rice, A. C. Ince-Cushman, M. L. Reinke, Y. Podpaly, M. J. Greenwald, B. LaBombard and E. S. Marmar, Plasma Phys. Control. Fusion {\bf 50}, 124042 (2008).

\bibitem{Diamond09}
P. H. Diamond, C. J. McDevitt, O. D. Gurcan, T. S. Hahm, W. X. Wang, E. S. Yoon, I. Holod, Z. Lin, V. Naulin and R. Singh, Nucl. Fusion {\bf 49}, 045002 (2009).

\bibitem{Diamond2013}
P. H. Diamond, Y. Kosuga, O. D. Gurcan, C. J. McDevitt, T. S. Hahm, N. Fedorczak, J. E. Rice, W. X. Wang, S. Ku, J. M. Kwon, G. Dif-Pradalier, J. Abiteboul, L. Wang,
W. H. Ko, Y. J. Shi, K. Ida, W. Solomon, H. Jhang, S. S. Kim, S. Yi, S. H. Ko, Y. Sarazin, R. Singh and C. S. Chang, Nucl. Fusion {\bf 53}, 104019 (2013).

\bibitem{Garbet PoP2013}
X. Garbet, D. Esteve, Y. Sarazin, J. Abiteboul, C. Bourdelle, G. Dif-Pradalier, P. Ghendrih, V. Grandgirard, G.
Latu, and A. Smolyakov, Phys. Plasmas {\bf 20}, 072502 (2013).

\bibitem{Lu Wang PRL2013}
L. Wang and P. H. Diamond, Phys. Rev. Lett. {\bf 110}, 265006 (2013).

\bibitem{Labit11}
B. Labit, C. Theiler, A. Fasoli, I. Furno, and P. Ricci, Phys. Plasmas {\bf 18}, 032308 (2011).

\bibitem{Terry}
J. L. Terry, S. J. Zweben, K. Hallatschek, B. LaBombard, R. J. Maqueda, B. Bai, C. J. Boswell, M. Greenwald, D. Kopon, W. M. Nevins, C. S. Pitcher, B. N. Rogers, D. P. Stotler and X. Q. Xu, Phys. Plasmas {\bf 10}, 1739 (2003).

\bibitem{Zweben}
S. J. Zweben, R. J. Maqueda, D. P. Stotler, A. Keesee, J. Boedo, C. E. Bush, S. M. Kaye, B. LeBlanc, J. L. Lowrance, V. J. Mastrocola, R. Maingi, N. Nishino, G. Renda, D. W. Swain, J. B. Wilgen and the NSTX Team, Nucl. Fusion {\bf 44}, 134 (2004).

\bibitem{Grulke}
O. Grulke, J. L. Terry, B. LaBombard and S. J. Zweben, Phys. Plasmas {\bf 13}, 012306 (2006).

\bibitem{Agositni}
M. Agostini, S. J. Zweben, R. Cavazzana, P. Scarin, G. Serianni, R. J. Maqueda and D. P. Stotler, Phys. Plasmas {\bf 14}, 102305 (2007).

\bibitem{Xu09}
G. S. Xu, V. Naulin, W. Fundamenski, C. Hidalgo, J. A. Alonso, C. Silva, B. Goncalves, A. H. Nielsen, J. Juul Rasmussen, S. I. Krasheninnikov, B. N. Wan, M. Stamp and JET EFDA Contributors, Nucl. Fusion {\bf 49}, 092002 (2009).

\bibitem{Chen10}
J. Cheng, L. W. Yan, W. Y. Hong, K. J. Zhao, T. Lan, J. Qian, A. D. Liu, H. L. Zhao, Y. Liu, Q. W. Yang, J. Q. Dong, X. R. Duan and Y. Liu, Plasma Phys. Control. Fusion {\bf 52}, 055003 (2010).

\bibitem{Maqueda}
R. J. Maqueda, D. P. Stotler and the NSTX Teama, Nucl. Fusion {\bf 50}, 075002 (2010).

\bibitem{Rudakov}
D. L. Rudakov, J. A. Boedo, R. A. Moyer, S. Krasheninnikov, A. W. Leonard, M. A. Mahdavi, G. R. McKee, G. D. Porter, P. C. Stangeby, J. G. Watkins, W. P. West, D. G. Whyte and G. Antar, Plasma Phys. Control. Fusion {\bf 44}, 717 (2002).

\bibitem{Boedo03}
J. A. Boedo, D. L. Rudakov, R. A. Moyer, G. R. McKee, R. J. Colchin, M. J. Schaffer, P. G. Stangeby, W. P. West, S. L. Allen, T. E. Evans, R. J. Fonck, E. M. Hollmann, S. Krasheninnikov, A. W. Leonard, W. Nevins, M. A. Mahdavi, G. D. Porter, G. R. Tynan, D. G. Whyte and X. Xu, Phys. Plasmas {\bf 10}, 1670 (2003).

\bibitem{Fundamenski}
W. Fundamenski, W. Sailer and JET EFDA contributors, Plasma Phys. Control. Fusion {\bf 46}, 233 (2004).

\bibitem{Endler}
M. Endler, I. Garcia-Cortes, C. Hidalgo, G. F. Matthews, ASDEX Team and JET Team, Plasma Phys. Control. Fusion {\bf 47}, 219 (2005).

\bibitem{Boedo}
J. A. Boedo, D. L. Rudakov, E. Hollmann, D. S. Gray, K. H. Burrell, R. A. Moyer, G. R. McKee, R. Fonck, P. C. Stangeby, T. E. Evans, P. B. Snyder, A. W. Leonard, M. A. Mahdavi, M. J. Schaffer, W. P. West, M. E. Fenstermacher, M. Groth, S. L. Allen, C. Lasnier, G. D. Porter, N. S. Wolf, R. J. Colchin, L. Zeng, G. Wang, J. G. Watkins, T. Takahashi and The DIII-D Team, Phys. Plasmas {\bf 12}, 072516 (2005).

\bibitem{Kirk}
A. Kirk, N. Ben Ayed, G. Counsell, B. Dudson, T. Eich, A. Herrmann, B. Koch, R. Martin, A. Meakins, S. Saarelma, R. Scannell, S. Tallents, M. Walsh, H. R. Wilson, and the MAST Team, Plasma Phys. Control. Fusion {\bf 48}, B433 (2006).

\bibitem{Herrmann}
A. Herrmann, A. Kirk, A. Schmid, B. Koch, M. Laux, M. Maraschek, H. W. Mueller, J. Neuhauser, V. Rohde, M. Tsalas, E. Wolfrum, ASDEX Upgrade Team, J. Nucl. Mater. {\bf 363-365}, 528 (2007).

\bibitem{Schmid}
A. Schmid, A. Herrmann, H. W. Muller and the ASDEX Upgrade Team, Plasma Phys. Control. Fusion {\bf 50}, 045007 (2008).

\bibitem{Dudson}
B. D. Dudson, N. Ben Ayed, A. Kirk, H. R. Wilson, G. Counsell, X. Xu, M. Umansky, P. B. Snyder, B. Lloyd, and the MAST Team, Plasma Phys. Control. Fusion {\bf 50}, 124012 (2008).

\bibitem{Silva}
C. Silva, W. Fundamenski, A. Alonso, B. Goncalves, C. Hidalgo, M. A. Pedrosa, R. A. Pitts, M. Stemp, and JET EFDA Contributors, Plasma Phys. Controlled Fusion {\bf 51}, 105001 (2009).

\bibitem{Hasegawa and Mima}
A. Hasegawa and K. Mima, Phys. Fluids {\bf 21}, 87 (1978).

\bibitem{Gurcan2010}
O. D. Gurcan, P. H. Diamond, P. Hennequin, C. J. McDevitt, X. Garbet, and C. Bourdelle, Phys. Plasmas {\bf 17}, 112309 (2010).

\bibitem{Gurcan2006}
O. D. Gurcan, P. H. Diamond and T. S. Hahm, Phys. Plasmas {\bf 13}, 052306 (2006)

\bibitem{Diamond book}
P. H. Diamond, S. I. Itoh, and K. Itoh, \emph{Modern Plasma Physics Vol.~1: Physical Kinetics of Turbulence Plasmas} (Cambridge University Press, 2010).

\bibitem{Muller 2011}
H. W. Muller, J. Adamek, R. Cavazzana, G. D. Conway, C. Fuchs, J. P. Gunn, A. Herrmann, J. Horacek, C. Ionita, A. Kallenbach, M. Kocan, M. Maraschek, C. Maszl, F. Mehlmann, B. Nold, M. Peterka, V. Rohde, J. Schweinzer, R. Schrittwieser, N. Vianello, E. Wolfrum, M. Zuin and the ASDEX Upgrade Team, Nucl. Fusion {\bf 51}, 073023 (2011).

\bibitem{ManheimerandDupree68}
W. M. Manheimer and T. H. Dupree, Phys. Fluids {\bf 11}, 2709 (1968).

\bibitem{Zhao2012}
L. Zhao and P. H. Diamond, Phys. Plasmas {\bf 19}, 082309 (2012).

\end{thebibliography}
\end{document}